\documentclass[twocolumn,letterpaper,aps,prc,showpacs,nofootinbib,floatfix]{revtex4-2}
\usepackage{multirow}
\usepackage{rotating}
\usepackage{notes2bib}
\usepackage{lineno}
\usepackage{longtable}  
\usepackage{graphicx}
\usepackage{amsmath,amssymb,bbold,bm}
\usepackage{epstopdf}
\usepackage{xcolor}
\usepackage{lipsum}
\usepackage{hyperref}
\usepackage{url}
\usepackage{xcolor}

\def\nonubb{$0\nu\beta\beta$}
\def\twonubb{$2\nu\beta\beta$}

\begin{document}

\title{Shake-up and shake-off effects in neutrinoless double-$\beta$ decay}
\author{J.\,A.~Detwiler}
\author{R.\,G.\,H.~Robertson}
\affiliation{Center for Experimental Nuclear Physics and Astrophysics, and Department of Physics, University of Washington, Seattle, Washington 98195, USA}
\date{\today}

\begin{abstract}
We revisit the role of shake-up and shake-off effects in neutrinoless double-$\beta$ decay, following earlier work by Drukarev {\it et al.}~\cite{Drukarev:2016dpp}.
We find in agreement with Drukarev {\it et al.}~that the $Q$ value of the decay is reduced by any atomic excitation in the final-state but not by the difference in atomic binding energy of the initial and final state atoms, as was recently suggested by Mei and Wei~\cite{Mei:2022mig, Mei:2022kxo}.
We discuss how the absorption of subsequent atomic de-excitation ejecta shifts the neutrinoless double-$\beta$ decay peak back to its nominal value.
We propose an {\it in situ} experimental verification of these ideas, and discuss the impact of shake-up and shake-off on two-neutrino double-$\beta$ decay spectral shapes. 
\end{abstract}


\maketitle

An intense global effort is underway to search for the asymmetric creation of matter without antimatter in the form of neutrinoless double-$\beta$ (\nonubb) decay~\cite{Agostini:2022zub}. 
The observation of this process would lend credence to grand unification theories and models of the cosmic matter imbalance that predict the neutrino to be a Majorana particle and mediate the decay.
At a deeper level, the existence of \nonubb\ decay would prove that $B-L$, which is accidentally conserved in the standard model, is not a conserved quantity in nature.
It would also prove that a mechanism exists to transform neutrinos into antineutrinos, strongly indicating that neutrino masses are governed by new physics.

Experiments aiming to detect \nonubb\ decay must distinguish it from the standard model process in which two neutrinos are emitted (\twonubb\ decay).
Neutrinos carry away a significant fraction of the available kinetic energy, resulting in ``missing energy'' in a calorimetric measurement.
In \nonubb\ decay, all of the available energy is effectively shared by the two emitted $\beta$'s, resulting in a monoenergetic peak at the $Q$ value of the process.
Knowledge of the true $Q$ value is thus critical to \nonubb\ decay searches.

$Q$ values are derived from precise measurements of neutral atomic masses of double-$\beta$ decay isotopes and their daughters~\cite{Wang:2021xhn}. 
Recently, Mei and Wei~\cite{Mei:2022mig, Mei:2022kxo} claimed that atomic effects shift the $Q$ value, and thus the observable \nonubb\ peak, downward by many keV from its usual assumed value.
If true, this would require a re-analysis of nearly all \nonubb\ decay searches to date.

The role of the atomic shell in $\beta\beta$ decay has been discussed before by Drukarev, Amusia, and Chernysheva~\cite{Drukarev:2016dpp}.
In this Letter we review how atomic phenomena affect \nonubb\ decay, and show that while $Q$ can indeed be shifted by atomic affects, namely by shake-up and shake-off, the shift is not the one proposed by Mei and Wei.
We assess how the impacts of shake-up and shake-off on a particular search depend on the experimental configuration, and find that no recent or proposed search has been sensitive to these effects.
We further propose a method to extract the position of $Q$ from \twonubb\ decay {\it in situ}.
We also sketch the impact of shake-up and shake-off on the \twonubb\ spectrum, and provide some guidance for accommodating their presence in future experimental efforts.

{\it Atomic effects in \nonubb\ decay.}
In \nonubb\ decay, an atomic nucleus hypothetically transforms into its second neighbor on the periodic table of nuclides, simultaneously emitting two new electrons and no antineutrinos. 
In all experiments performed to date, the initial-state nucleus is deployed as a neutral atom in a detection apparatus capable of measuring the summed kinetic energy of the two electrons, which is typically several MeV.

Despite being a nuclear process, the energy available in the decay is shared among not only the two new electrons and the final state nucleus, but also the final state atomic electrons. 
Since the two new electrons are relativistic, the sudden approximation is valid and the final state atomic electron configuration $|\psi'\rangle$ is essentially (to within thermal fluctuations) that of the initial atomic ground state $|\phi_0\rangle$, which can be written as a superposition of the atomic electron eigenstates of the doubly ionized final state atom:
\begin{equation}
|\psi'\rangle = \sum_n  \langle \phi'^{++}_n |\phi_0\rangle |\phi'^{++}_n\rangle,
\label{eq:superpos}
\end{equation}
where $|\phi^s_n\rangle$ denotes the $n$th atomic energy eigenstate of species/ionization state $s$, with excitation energy $E^s_n$. 
Unprimed and primed species refer to the initial and final states, respectively.
When any particular decay occurs, the final-state atomic electrons are found to be in state $|\phi'^{++}_n\rangle$ with probability $| \langle \phi'^{++}_n |\phi_0\rangle |^2$.
For this final state, the $Q$ value is:
\begin{equation}
Q_n = M\!\left(^{A}_{Z}X\right) c^2 - 2m_e c^2 - M\!\left(^{~~~A}_{Z+2}X'^{++}\right) c^2 - E'^{++}_{n},
\end{equation}
where $^{A}_{Z}X$ refers to atomic species $X$ with $A$ nucleons including $Z$ protons, and $M(X)$ is the mass of the corresponding atomic ground state.
When $n>0$, the atomic excitation is referred to as ``shake-up.''
We can include in $n$ the continuum of states in which one or more of the atomic electrons are unbound following the decay [the sum in Eq.~(\ref{eq:superpos}) becomes an integral, and $E^s_n$ will include the kinetic energies of the unbound electrons]; this situation is referred to as ``shake-off.''

The mass of the ground state of the doubly ionized final-state ion represented by $M(^{~~~A}_{Z+2}X'^{++})$ deviates from the mass of the corresponding neutral atom $M(^{~~~A}_{Z+2}X')$ by $2 m_e$ minus the binding energies of the outermost atomic electrons (i.e., the minimum binding energies) of the neutral $(b_{min})$ and singly ionized $(b^+_{min})$ atom. 
We thus find that when the atomic electrons in the final state are in atomic excitation level $n$, $Q_n$ is given by
\begin{equation}
Q_n = M(^{A}_{Z}X) c^2 - M(^{~~~A}_{Z+2}X') c^2 -  b'_{min} - b'^+_{min} - E'^{++}_{n}.
\label{eq:Qf}
\end{equation}
The neutral atomic mass difference $M(^{A}_{Z}X) c^2 - M(^{~~~A}_{Z+2}X') c^2$ is precisely that which is measured with high accuracy in atomic trapping experiments.
Thus the $Q$ value is indeed shifted due to atomic effects.
However, in contrast to~\cite{Mei:2022mig,Mei:2022kxo} (and in agreement with~\cite{Drukarev:2016dpp}), we do not find that that shift is given by the difference in the total atomic binding energies of the ground states of the initial- and final-state atoms.
It is only shifted by $b_{min} + b^+_{min}$, which is $\lesssim$~100~eV for $\beta\beta$ nuclei of experimental interest~\cite{xrdb:2009}, plus any energy that goes into the {\it excitation} of the final state atom.
Mei and Wei reach an erroneous expression for $Q$ because they begin with an equation that assumes bare nuclei rather than atoms in the initial and final states.
They also do not account for how the released energy is shared between the emitted $\beta$'s and the atomic electrons.

Additionally, as opposed to~\cite{Mei:2022mig,Mei:2022kxo} and as suggested in the summary of~\cite{Drukarev:2016dpp}, we do not find a single $Q$ value for double-$\beta$ decay, but one for each possible atomic excitation state $n$.
The rate for the occurrence of each is governed by the probabilities $| \langle \phi'^{++}_n |\phi_0\rangle |^2$. 
What is computed in \cite{Drukarev:2016dpp} is the energy-weighted mean of the $Q_n$ for shake-off states, but is otherwise consistent with our formulas. 
For scale, Drukarev {\it et al.}~find an average $\langle E'^{++}_{n} \rangle = 300$--$500$~eV for $^{76}$Ge and $^{136}$Xe.

It is worth noting that modern high-precision calculations of \nonubb-decay phase-space factors~\cite{Kotila:2012zza,Stefanik15,Neacsu18} appear to ignore these atomic effects.
Since the \nonubb\ phase space factor scales roughly as $Q_{\beta\beta}^{7}$~\cite{Haxton:1984ggj}, an excitation on the order of the $K$-shell binding energy would suppress the phase space factor by as much as $\approx$~1\%.
However, the average shift computed by Drukarev {\it et al.} corresponds to a suppression by only $\approx$~0.1\%.

The presence of additional energy sinks, such as final state interactions or molecular excitations, can be accommodated in a similar manner. 
The $Q$ value for \twonubb\ decay is identical modulo the masses of the emitted neutrinos.
A similar form is found, e.g., using a full $N$-body treatment of the $\beta$ decay of molecular tritium~\cite{Bodine:2015sma}.
An additional shift is contributed by the difference in molecular/chemical bindings of the parent and daughter species,
but shifts are small on the scale of current experimental energy resolutions.
The case of tritium embedded in silicon and decaying to $^3$He is possibly an extreme example, and is only 10 eV~\cite{Redondo:1989wk}.

{\it Experimental Considerations.}
$Q_n$ corresponds to the maximum kinetic energy that the two emitted electrons can share. 
Although the nucleus can also recoil and carry away some of the energy, its maximum kinetic energy is on the order of tens of eV, which like $b_{min} + b^+_{min}$ is much smaller than the FWHM of even the highest-resolution \nonubb-decay detectors (e.g., \cite{Majorana:2022udl}). 
In principle, energy could also be carried away by (inner) bremsstrahlung photons or conversion electrons etc.~generated due to final-state interactions during the process. 
But for events without such ejecta,  the summed kinetic energy of the two electrons will be equivalent to $Q_n$, which to within the measurement uncertainty (i.e., neglecting $b^{(}\!'^{)}\!_{min}$) is given by the neutral atomic mass difference shifted by the final state atomic excitation energy $E'^{++}_{n}$.

Among the possible final states is that for $n=0$, i.e., the final-state atomic electrons end up in their ground state. 
In this case $E'^{++}_{0} = 0$ and $Q_n$ has its maximum value, given effectively by the neutral atomic mass difference. 

When $n > 0$, the \nonubb-decay event is followed immediately by the de-excitation of the final state atom.
Atomic de-excitations in charged ions typically occur at the ns scale or faster~\cite{atoms10020046}, which is faster than the timescales for collection of photons ($>$ns), ionization ($\approx$$\mu$s), or phonons ($>$ms) in double-$\beta$ decay experiments.
The resulting ejecta are low-energy x rays and Auger electrons, which have absorption lengths much smaller than the two high-energy $\beta$'s emitted in the decay. 
For experiments with the $\beta\beta$ isotope embedded in the detector's active region, these atomic de-excitation ejecta will generate signals and contribute to the collected event energy. 
Extremely low-energy ejecta may have their energy deposition quenched relative to higher-energy depositions by $\beta$'s due to atomic physics that modifies scattering cross sections at the lowest incident energies~\cite{DAMIC-M:2022xtp}. 
However this only occurs at energies far below the best-available energy resolutions and can be neglected.

Most modern detectors indeed use embedded sources, and so in these detectors the total collected energy sums to the neutral atomic mass difference. 
A notable exception is the NEMO/SuperNEMO design~\cite{NEMO-3:2015jgm,Piquemal:2006cd}, which uses source foils. 
Atomic de-excitations indeed can be absorbed in the foils and go undetected. 
However in these detectors the energy resolution near $Q_n$ is much greater than the typical value of $E'^{++}_{n}$. 

Thus, for all double-$\beta$ decay experiments performed to date, to within the experimental energy resolution, the \nonubb-decay spectrum should appear as a single, monoenergetic peak at $Q_{\beta\beta}$, given by
\begin{equation}
\label{eq:qbb}
Q_{\beta\beta} \approx M\!\left(^{A}_{Z}X\right) c^2 - M\!\left(^{~~~A}_{Z+2}X'\right) c^2
\end{equation}
It is possible that future experiments could have differing sensitivity to atomic de-excitations.
Thus the validity of the approximate equality in Eq.~(\ref{eq:qbb}) must be evaluated on a case-by-case basis, considering the identities, energies, and timing of the de-excitation quanta relative to a particular detector's features and capabilities.

{\it In-Situ Experimental Determination of $Q_{\beta\beta}$.}
In single $\beta$-decay experiments it is possible to determine the $Q$ value in a fit to spectral data, for example using a Kurie plot.
Allowed transitions emit $\beta$'s with kinetic energy $E$ and momentum $p = \sqrt{E^2 + 2mE}$ with differential rate
\begin{equation}
\frac{d \Gamma_{\beta}}{dE} = A \, F(Z, p) \, (Q - E)^2,
\end{equation}
where $F(Z,p)$ is the Fermi function, $Z$ is the charge of the daughter nucleus, and $A$ is a constant.
By fitting a plot of $\sqrt{\frac{d \Gamma_{\beta}}{dE} / A \, F(Z, p)}$ vs.~$E$ to a straight line, one obtains $Q$ as the $x$-intercept of the line.

In modern experiments, fits are made directly to the full spectral shape, incorporating corrections to the theoretical expression for the differential rate, as well as experimental artifacts and systematic considerations.
A state-of-the-art fit was recently performed by the KATRIN experiment for the decay of tritium~\cite{KATRIN:2021uub}, giving a $Q$ value of ($18575.20 \pm 0.60$)~eV, consistent with the  
$^3$H--$^3$He neutral atomic-mass difference~\cite{Myers:2015lca} corrected for the molecular bindings and ionization~\cite{Otten:2008zz,KATRIN:2019gru}, giving ($18575.72 \pm 0.07$)~eV.
Applying the technique of \cite{Mei:2022mig} to tritium decay, a $Q$-value shift of 48.6~eV would be expected, which is highly inconsistent with the KATRIN data.

One may perform similar fits to experimentally verify Eq.~(\ref{eq:qbb}) for \nonubb\ decay {\it in situ}.
In fact, the Kurie plot technique can be used also in the case of \twonubb\ decay, since
the differential rate vs summed energy $E_{\Sigma}$ of the two $\beta$s emitted in \twonubb\ decay is 
well approximated by a function of $E_{\Sigma}$ times $(Q_{\beta\beta} -E_{\Sigma})^5$~\cite{Haxton:1984ggj}:
\begin{equation}
\label{eq:2nbbrate}
\frac{d\Gamma_{2\nu\beta\beta}}{dE_{\Sigma}} \approx A(E_\Sigma) (Q_{\beta\beta} - E_{\Sigma})^5.
\end{equation}
Thus a straight-line fit of $\left( \frac{d\Gamma_{2\nu\beta\beta}}{dE_{\Sigma}} / A(E_\Sigma) \right)^{1/5}$ vs. $E_\Sigma$ would similarly yield $Q_{\beta\beta}$ as the $x$ intercept of the line.

\begin{figure}[htbp]
   \centering
   \includegraphics[width=0.5\textwidth]{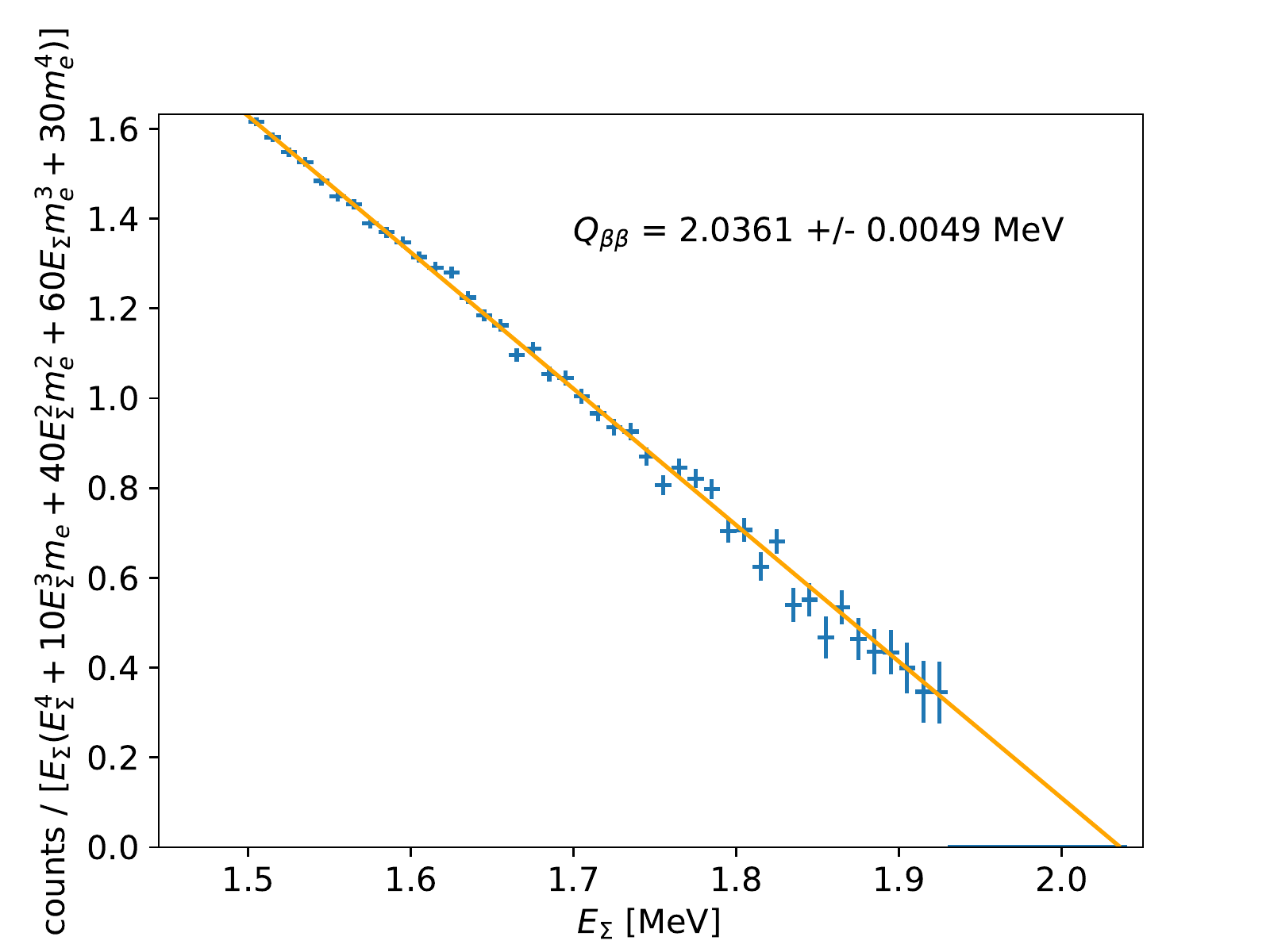}
   \caption{$\beta\beta$ Kurie plot for $10^6$ simulated $^{76}$Ge double-$\beta$ decay events (blue crosses). The linear fit (orange line) is performed within 500~keV of $Q_{\beta\beta}$. The best-fit value of $Q_{\beta\beta}$ and its statistical uncertainty are given. }
   \label{fig:bbkurie}
\end{figure}

However, due to the $(Q_{\beta\beta} - E_{\Sigma})^5$ in Eq.~(\ref{eq:2nbbrate}), the statistics near $Q_{\beta\beta}$ are very low.
Most of the constraint on $Q_{\beta\beta}$ in the fit comes from counts from much lower relative energies than for the case of single-$\beta$ decay.
We performed a Monte Carlo simulation to estimate the statistical precision of such a fit.
For $A(E_\Sigma)$ we used the polynomial approximation in~\cite{Haxton:1984ggj} (originally from~\cite{Primakoff:1959chj}), and then generated a simulated spectrum following Eq.~(\ref{eq:2nbbrate}) with $10^6$ counts, which is similar to the total number of counts collected in recent and upcoming experiments, such as KamLAND-Zen~\cite{KamLAND-Zen:2022tow} and LEGEND-200~\cite{LEGEND:2021bnm}.
For definiteness, we used $Q_{\beta\beta}$~=~2039~keV for $^{76}$Ge; no energy resolution smearing was applied to the simulated energies.
We then plotted the $\beta\beta$ version of the Kurie plot for the simulated data, shown in Fig.~\ref{fig:bbkurie}.
As expected, the data are very sparse near $Q_{\beta\beta}$, with very few counts within a hundred keV or so at these statistics.
A linear fit was performed starting from $Q_{\beta\beta}\!-\!500$~keV. The fit value of $Q_{\beta\beta}$ was found to have an uncertainty on the order of 5~keV, on the same order of the size of the hypothetical shifts computed in~\cite{Mei:2022mig, Mei:2022kxo}. We note, however, that references~\cite{Mei:2022mig, Mei:2022kxo} appear to erroneously use the sum of binding energies for individual atomic electrons in the neutral atom to approximate the total atomic binding energy, thus their proposed shifts are likely strongly underestimated.

We thus conclude that current experiments may be unable to validate Eq.~(\ref{eq:qbb}) relative to the hypothesis in~\cite{Mei:2022mig, Mei:2022kxo} {\it in situ}. 
However, the future experiment CUPID~\cite{cupid-cdr} will perform a \nonubb-decay search with $^{100}$Mo, whose \twonubb-decay rate is the highest known.
CUPID uses calorimetric crystals for which Eq.~(\ref{eq:qbb}) applies, and prototype detectors have exhibited very good energy resolution.
Containing more than 250~kg of $^{100}$Mo, CUPID will collect $>$10$^8$ \twonubb\ decay events per year.
This gives a factor $>$~10 improvement in statistical uncertainty over current measurements, providing sensitivity to shifts on the order of $\langle E'^{++}_{n} \rangle$.
Thus CUPID should be able to perform a meaningful experimental determination of $Q_{\beta\beta}$. 
In a ten year run, CUPIDs statistical precision would reach $\approx$~0.2 keV, on the same order as the 0.17~keV uncertainty in the $^{100}$Mo $Q$-value determined in Penning trap measurements~\cite{Rahaman:2007ng}.

{\it Impact on \twonubb-Decay Spectra. }
We remark that shake-up and shake-off have to date been neglected in state-of-the-art computations of \twonubb-decay spectra (e.g., \cite{Kotila:2012zza}).
To first order, the presence of a nonzero $E'^{++}_{n}$ can be treated with a simple squeezing of the spectrum that keeps the rate constant:
\begin{equation}
\frac{d\Gamma_{2\nu\beta\beta}(E_{\Sigma})}{dE_{\Sigma}} = \frac{1}{1-\frac{E'^{++}_{n}}{Q_{\beta\beta}}} \frac{d\Gamma_{2\nu\beta\beta}(\tilde{E}_{\Sigma})}{d\tilde{E}_{\Sigma}}
\end{equation}
where $\tilde{E}_{\Sigma}$ is the energy in the spectrum extending all the way to $M(^{A}_{Z}X) c^2 - M(^{~~~A}_{Z+2}X') c^2$, i.e.~$E_{\Sigma} = \tilde{E}_{\Sigma} \left(1-\frac{E'^{++}_{n}}{Q_{\beta\beta}} \right)$.
It should be understood that every event from a spectrum suppressed by $E'^{++}_{n}$ will be accompanied by atomic de-excitations.
If the energy depositions from those de-excitations are collected, the associated measured energy will be $E_{\Sigma} + E'^{++}_{n}$.
In this case, the resulting differences in the lowest-energy region of the spectrum are expected to be negligible compared to the contributions from background processes in that region.

The presence of $E'^{++}_{n}$ will additionally result in higher-order distortions to the shape of the spectrum, but conceivable experiments are likely not sensitive to those distortions.
However, although $E'^{++}_{n}/Q_{\beta\beta}$ is of order 0.1\%, since the \twonubb\ decay rate scales roughly as $Q_{\beta\beta}^{11}$~\cite{Haxton:1984ggj}, atomic excitations can lead to non-negligible suppression of the total rate, on the scale of a percent or so.
This is on the same order as the state-of-the-art experimental uncertainty in measurement of the decay rate~\cite{EXO-200:2013xfn}.
However, experiments require only a probability density distribution for $\frac{d\Gamma_{2\nu\beta\beta}(E_{\Sigma})}{dE_{\Sigma}}$, for which the absolute normalization is irrelevant.
And the absolute normalization is also only relevant in combination with nuclear matrix elements, which have much larger uncertainties.
Still, these calculations should be repeated with the appropriate atomic physics included should higher accuracy in the absolute normalization or the spectral shapes be required.

{\it Conclusions.}
The $Q$ value determines the energy at which \nonubb-decay experiments search for a hypothetical monoenergetic peak.
Its position depends not only on the nuclear transition but atomic effects such as shake-up and shake-off as well, as pointed out in~\cite{Drukarev:2016dpp}.
We find that the \nonubb-decay $Q$ value has not one but a spectrum of values, given to a very good approximation by the difference in the neutral atomic masses of the initial and final state species, minus the spectrum of atomic excitations of the final-state doubly charged ion.

In all experiments known to the authors to date, those atomic excitation energies are either below the experimental energy resolution, or are subsequently collected and contribute to event energies.
Experiments for which these conditions apply should indeed expect to see a monoenergetic peak at $Q_{\beta\beta} \approx M(^{A}_{Z}X) c^2 - M(^{~~~A}_{Z+2}X') c^2$.
The suggestion in~\cite{Mei:2022mig} and \cite{Mei:2022kxo} that $Q_{\beta\beta}$ is shifted below this value by an amount equal to the difference in atomic binding energies is found to be in error.
Such a hypothesis could be probed in future double-$\beta$ decay experiments, but it is already ruled out with high confidence by single-$\beta$-decay spectral measurements.

Shake-up and shake-off have an impact on \nonubb-decay phase space factors as well as the spectrum of \twonubb\ decay, but their presence is essentially negligible there as well.
Future experiments with higher statistics should evaluate their sensitivity to shake-up and shake-off, especially if new breakthroughs are made in energy resolution or event timing and reconstruction that make this physics non-negligible.

{\it Acknowledgments.}
This work was performed with support from the U.S.~Department of Energy, Office of Science, Office of Nuclear Physics under grant number DE-FG02-97ER41020.
The authors would like to thank J.~Men\'{e}ndez and members of the LEGEND  and SuperNEMO Collaborations for useful discussions.

\bibliography{qshift_paper}
\end{document}